# Cyber-Physical Queueing-Network Model for Risk Management in Next-Generation Emergency Response Systems

**Mengran Xue[1] and Sandip Roy**

*Abstract –* Queueing-network models are developed for enhanced and next-generation 911 (E911 and NG911) systems, which capture both their cyber- components (communications, data-processing) and physical-world- elements (call takers, vehicle dispatch).  The models encompass both the call-processing and dispatch functions of 911, and can represent the interdependencies between multiple PSAPs enabled by NG911.  An instantiation of the model for a future NG911 system for Charlotte, North Carolina, is developed and used to assess performance metrics.  Representation of cyber- threats (e.g. Distributed Denial-of-Service attacks) within the queueing-network model is undertaken.  Based on these representations, the model is used for analysis of holistic threat impacts, as a step toward risk and vulnerability assessment for future emergency response systems.

## 1. Introduction

Next Generation 911 (NG911) is an initiative to update the United States' emergency communications infrastructure to an Internet Protocol (IP)-based system [1-5].  This update to an IP-based system holds promise to improve functionality of the 911 system, primarily in two ways.  First, NG911 will allow the public seamless access to 911 services via multiple modes, encompassing traditional telephony, text messaging, Voice-over-IP (VoIP), and image and video transmission.  Second, NG911 systems will provide the facilities responsible for emergency communications (the public safety access points, or PSAPs) with the interconnectivity necessary for improved locationing, management of call overload, and resilience to failure.  Given the essential role of 911 in maintaining public safety, the new functions supported by NG911 should translate to concrete health, safety, and socioeconomic benefits.  Thus, the NG911 modernization initiative has garnered considerable interest at the federal and state level in the United States, and the necessary technologies are gradually being put in place – either through targeted updates to the Enhanced 911 (E911) systems that are currently predominant  (e.g. to allow text messaging or calls from cellphone-based apps), or through full-scale piloting of NG911. Parallel efforts to modernize emergency communications systems are also underway in Europe and Asia [51].

The deployment and operation of NG911 systems is bringing forth a number of technological challenges, which broadly stem from the specialized time- and safety- critical nature of 911 systems, and the complex coordination of communication systems, physical-world assets, and human operators required for emergency service. One critical challenge is the management of cyber risks, encompassing malfeasance of various sorts as well as inadvertent failures and errors [6,7]. Cyber- risks are already a concern in today's E911 systems, as evidenced by several high-profile events such as a denial-of-service attack causing a multi-state outage, repeated outages due to failures in Washington State's system, and a cyber- attack on emergency communication systems in Phoenix, AZ. Unfortunately, cyber risks are expected to be many-fold amplified in NG911 systems, for several reasons. First, as an IP-based system, the NG911 emergency communication infrastructure will be tied to the Internet-at-large (in sharp contrast with current systems) and will allow for multi-modal access, leaving the system open to a diverse array of known and unexpected threats. In cyber-security parlance, the threat surface will be greatly increased, and hence the likelihood of both known and unknown threats will be increased. Second, NG911 proposes to increase and automate the coordination of PSAPs via the Emergency Services IP Network (ESINet) construct, which forms the network backbone of the new system. While this additional coordination potentially can bring forth benefits in terms of resilience, it also means that threats may create propagative impacts across wide areas, greatly increasing their potential impact or consequence. Third, the piecemeal deployment of NG911 means that interoperability issues are inevitable, also creating vulnerabilities and increasing threat consequences. In sum, both the likelihood and the consequences of cyber threats, and hence risk, is likely to be significantly amplified in NG911 systems.

[1] The work described here was partially supported by the Department of Homeland Security through a Small Business Innovative Research contract to Achilles Heel Technologies Inc., and was partially conducted on the authors' own time. Please send correspondence to sroyeecs@gmail.com . We acknowledge communications with B. Moss, S. Warnick, and D. Grimsman on the topic of this work.

Authorities involved in planning NG911 have recognized the critical importance of managing cyber risks, both via appropriate system design and via persistent monitoring and response during regular operations. Specifically, the common architecture for NG911 proposed by the National Emergency Number Association (NENA) acknowledges cyber-security as a critical concern and examines in detail case scenarios of representative threats and their possible consequences [6]. Additionally, a working group led by the Department of Homeland Security's Office of Emergency Communications has developed a document (colloquially known as the "Cybersecurity Primer") which overviews the cyber risk management landscape as summarized above [7], and provides guidelines for organizations involved in NG911 including state and regional 911 offices and PSAPs. While these efforts have created promising frameworks for risk management, however, to date systematic methodologies and toolsets have not been developed for holistically assessing and designing against cyber- risks in planning NG911 systems, nor for providing real-time situational awareness and adaptation for cyber threats during operations. Such tools and methods for risk management would be directly usable by: 1) state and regional 911 offices and technology vendors to assess cyber threats prior to deployment of NG911 systems and to develop risk plans for operational systems; and 2) PSAP operators to gain situational awareness and understand mitigation strategies for actualized threats.

The purpose of this study is to introduce a modeling approach, which supports assessment and management of cyber-risks to NG911 systems. Specifically, our primary aim is to enable quantitative assessment of the holistic *consequences or impacts* of cyber threats on the overall performance of the emergency response system, as measured by the fidelity and delay in deploying assets (e.g. police cars, fire trucks, ambulances) to called-in emergencies. Our modeling approach recognizes that emergency response systems involve a tight intertwining of cyber (communication/computing), physical-world, and human assets. These assets each play a role in processing or serving emergency calls, which are discrete events that arise in a stochastic fashion. Importantly, successful service of a call depends on sequential processing by these varied assets, and further each processing step significantly contributes to the overall delay in responding to a call. Many of these elements are also subject to threats which may have propagative impacts on system performance. Given these characteristics, holistic performance modeling requires integrated modeling of cyber and physical processing elements – i.e., it requires a cyber-physical model. In addition, NG911 systems are distinguished by the networking of multiple emergency response systems (specifically PSAPs), and the adaptability and accessibility of these systems. Thus, NG911 systems also require network-theoretic models, which can capture the interplay among multiple PSAPs which are interconnected via the IP backbone, and also multi-model call-in access. Based on these characteristics, here we propose a queueing-network model for NG911 systems, which encompass both cyber and physical service elements. We further pursue representation of threat classes in the context of the queueing-network model. After defining the model, we undertake performance assessment of the nominal system, as well as simulation of specific threat events. We also discuss model parameterization in a preliminary way, via an illustrative example.

The research described here is part of an emerging research focus on risk management in critical infrastructure sectors which are becoming hyper-vulnerable to cyber threats, due to the integration of pervasively-networked cyber systems (e.g., agriculture, air transportation, power delivery). For these various infrastructure sectors, a wide array of methods and software tools are being developed which aim to provide system stakeholders with holistic real-time situational awareness and risk mitigation capabilities for threats that arise in the cyber- world. The particular models and methods for cyber- risk management naturally vary from one sector to the next, but have in common that they recognize the cyber- and physical- elements of the infrastructures, and aim to assess propagative impacts of cyber- events. In this sense, the work differs from both traditional cyber-security which is narrowly focused on cyber- exploits, and traditional risk-management work which does not encompass cyber processes and risks. Our research in this area has been focused on using dynamical network-models to undertake quantitative counterfactual or "what-if" analysis of threat impacts [8-25]. At its essence, our approach builds on classical domain-specific models to obtain dynamic network models, which capture both physical-world and cyber interconnectivities in the infrastructure. These models together with control- and data- theoretic approaches are used for accelerated analysis of holistic consequences of potential threats. These fast threat and vulnerability assessments allow identification and mapping of key risks, and in turn support risk planning and real-time threat monitoring. The research described here contributes to the dynamical network modeling approach, by introducing a new class of dynamical network models (specifically, queueing-network models) which represent stochastically-driven cyber-physical service processes such as the emergency response system. The described research is also part of an extensive research effort on modeling and control of cyber-physical systems, which has been driven by pervasive integration of communications and computing technologies in the engineered world, which encompasses infrastructures as well as other systems [35-50].

The remainder of the article is organized as follows. A brief technical review of 911 systems is given in Section 2. The queueing-network model is introduced in Section 3, and the model is instantiated for a case study in Section 4. Threat representation and threat-impact analysis using the model is discussion in Section 5. Finally, the implications of the model in risk assessment/management are briefly discussed, and future directions of work are listed, in Section 6.

## 2. Technical Review of 911 Systems: Architecture Description and Existing Models

Based on the overall organization of the 911 system, it is natural to decompose the system into two major functional parts: 1) connection and communication between the originating caller and the appropriate PSAP (often called *call processing*) and 2) dispatch of physical assets based on communication from PSAPs (called *dispatch*). Here, these two sequential parts of the 911 system are described from an operational perspective, and some relevant technical literature is also presented with a focus on existing mathematical models.

*Part 1: Interconnection between the originating caller and the proper PSAP*

The first crucial function of a 911 system is to make sure that the originating caller successfully connects to a PSAP, and in turn communicates with a call-taker. In legacy 911 systems (i.e. Basic 911 and E911), the communication between the caller and the PSAP is based on telephony (e.g. public switched telephone networks). These communication mechanisms traditionally only supported voice calls. Recently, some E911 systems have been extended to support limited "text-to-911" and perhaps cell-phone-app-based services, but not multimedia data. The primary difference between basic 911 and E911 is the automated provision of location data in E911.

As information technologies become increasingly advanced, the ways in which people are communicating with each other is changing. Data is no longer limited to voice only: text, images, video, and calls from VoIP are becoming more and more popular data formats in people's daily communications. This also has motivated the modernization of the current 911 system to a new architecture, which is called Next Generation 911 (NG911). The new technologies in NG911 are meant to allow PSAPs to process all types of emergency calls including non-voice (multimedia) messages. At the same time, the new technologies provide alternative means for communication and interconnection among PSAPs, which are IP-based rather than telephony-based. During the transition from the legacy 911 system to NG911, a key challenge in implementing these new technologies is how to standardize interfaces from different call services, as the proposed system will include mixtures of calls routed through wired and wireless telephone network, and through IP networks.

The National Emergency Number Association (NENA) proposed an IP-based architecture for NG911, namely NENA i3, which serves as one standard for a redesigned 911 system which uses IP-based technology [2]. Specifically, this IP-based architecture extends the traditional 911 service to a wider range of callers by 1) supporting multimedia data, 2) accepting calls from the Internet, and 3) allowing flexible call routing to proper PSAPs based on callers' geographical locations and PSAP/dispatch statuses. The core service in the NENA i3 architecture is an IP-based network called Emergency Service IP Network (ESInet), which provides the flexible routing and data transfer functions. One ESInet serves one or multiple PSAPs and is linked to other ESInets in a hierarchical structure. A unification of core NG911 services is provided through various interfaces in the ESInets. These core services are principally concerned with providing accurate and reliable call routing services to the appropriate PSAP and caller location and number information in a modern, geo-spatial format; we note that accurate locationing of VoIP calls and IP data transmissions represents a significant technical challenge. ESInets may still use the existing 911 Selective Routers (911SRs) to help with routing the data and information. During the transition period, some PSAPs may not have direct access to ESInets, in which case post-processing steps are required between different formats of data/signals. For example, PSAPs that do not support multimedia data transfer may need to send the voice information via a legacy PSAP gateway to the local ESInet. Figure 1 abstractly diagrams how data and information is transferred between various parties and the ESInets. The diagram highlights the regional-scale network structure of the IP-based services and PSAPs in NG911, which is foundational to our network-modeling efforts.

Since we are interested in developing predictive models for risk assessment, the state-of-the-art in modeling the caller-to-PSAP interconnection is also scoped. Many descriptive models are available for basic/enhanced 911 systems (e.g. [26]). Likewise, the call architecture for NG911 has been qualitatively modeled [1-5], although the descriptions are somewhat lacking in detail since deployment of NG911 is in its preliminary stages. To the best of our knowledge, complete mechanistic mathematical models have not been developed for 911 call processing (either basic/enhanced or next generation), although some statistical descriptions for performance features are available (e.g. for time/accuracy of locationing and total call-processing time) [27]. Also, many quantitative models are available for general communication systems, including telephony and IP-based networks, which potentially could be leveraged to assess 911 and NG911 [28]. Broadly, these quantitative models permit prediction of communication-system performance metrics (e.g., congestion, delay, quality-of-service), and could be used to represent the communication systems within 911 and NG911.

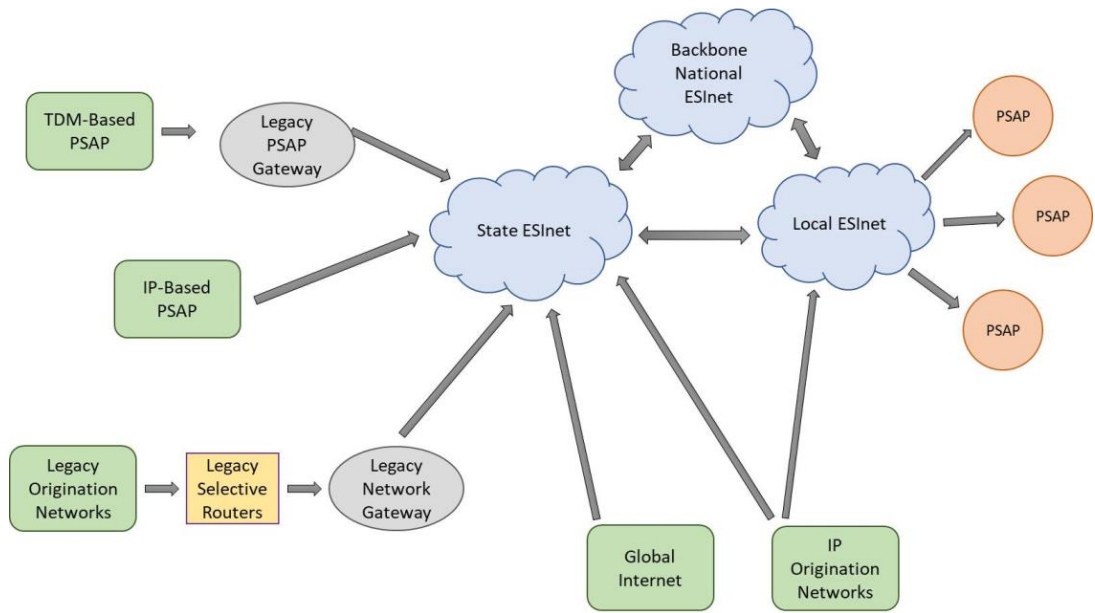

**Figure 1:** A high-level flowchart of communications in NG91 based on NENA i3 architecture.

*Part 2: Dispatch of physical assets based on communication from PSAPs*

The dispatch of resources (e.g., police, fire, ambulance) by PSAP personnel is a traditional component of emergency response systems. While the dispatch process has gradually evolved with new technological innovations (e.g. computer-aided dispatch technologies), it is principally defined by spatiotemporal physical-world processes (e.g., routing of emergency vehicles within the traffic network) which have remained relatively consistent over 50+ years. This consistency has allowed for the development of not only qualitative but quantitative models for the performance of emergency service dispatch. A primary modeling approach used for quantitative analysis is a queueing network paradigm, which was introduced by Larson and co-workers in the early 1970's [29] and has been enriched extensively since then (e.g. [30-32]). The modeling paradigm, known as the *hypercube queueing model*, captures the spatial and temporal aspects of the dispatch process and admits performance assessment through simulation as well as formal methods. More specifically, the modeling approach assumes that mobile resources or assets are distributed within a geographical area, and that emergency service needs arise according to a spatiotemporal stochastic process within that area. For each event, the nearest available salient resource is dispatched to the service location, with the time taken to reach the location being dependent on distance. The parameterization of the model from field data, as well as evaluation/validation of the model, have also been pursued extensively, and the model is known to provide reasonable statistical estimates of dispatch times in dense areas.

The qualitative and quantitative models for dispatch were developed for traditional 911 systems. Indeed, the models do not capture the call-processing part of emergency response, and hence only providing estimates of dispatch characteristics. For NG911, dispatch is expected to be largely unaltered compared to traditional 911. One possible alteration is a higher or changed frequency of cross-dispatch among PSAPs, since NG911 will likely incorporate automated re-routing among PSAPs depending on congestion and perhaps availability of dispatch assets. Models therefore may need to capture dispatch in adjacent geographical areas covered by different PSAPs; the hypercube queueing model can be naturally enhanced to cover such cross-dispatch.

## 3. Cyber-Physical Queueing-Network Model

Models for E911 and NG911 systems are presented, which can assist with cyber threat impact (consequence) assessment. To enable cyber-threat representation and consequence analysis, the models are designed to encompass both the call-processing and dispatch functions of 911 systems, and hence capture cyber- (telephony, IP-based communications) and physical- (call takers, dispatch) processes. In developing the models, we note that 911 communication and dispatch processes involve discrete arrivals events (e.g. service calls), requiring processing by discrete elements (e.g. routers, dispatched assets). These processes are not easily represented by the differential-equation models that are traditionally used to capture many infrastructure dynamic processes. However, they are at their essence networked state-space dynamics – i.e., they can be described by tracking terse states associated with discrete service components (statuses of call takers, availability of slots at routers/gateways, locations of dispatch assets), together with the stochastic arrival processes. Further, these components' states evolve in an interdependent way, due to the interdependence of call processing and dispatch (for all 911 systems) and interdependence among PSAPs (for NG911). Processes of these types can be represented using queueing network models, and indeed queueing models have been used to represent diverse communication processes [28] and separately emergency dispatch processes [29-32] as overviewed above. These queueing network models are easier to parametrize and computationally simpler than detailed (e.g. packet-level) simulation models, yet sufficiently rich that they can capture dynamic behaviors and interdependencies of the systems. Thus, a queueing-network modeling framework is appealing for E911 and NG911 systems.

The base module in our framework is a queueing-network model for a single PSAP in a basic or enhanced 911 system, which represents both client communication with the PSAP (call processing function) and the dispatch of response assets (dispatch function). Performance assessment of a basic or enhanced 911 system requires modeling both the call-processing and dispatch functions of a PSAP. Traditionally, delays in dispatch in congested urban areas have been a primary concern in 911 system operations, and hence the primacy of the modeling efforts has focused on representing the dispatch process. Indeed, a family of queueing-theoretic models has been developed for emergency dispatch in urban areas, starting from the fundamental work of Larson and co-workers on the hypercube queueing model [29]. These efforts represent the dispatch process using queueing or stochastic state-space model at various resolutions, under the assumption that the communication functions of the 911 system (including the needed telephony and the client-dispatcher communications) are entirely reliable and incur minimal delay. We have extended the hypercube-based models for dispatch to also represent the call-processing function of a PSAP, with the motivation that: 1) unreliability of call-processing/communications may be increasingly of concern in Internet-based next-generation systems and 2) threats to the communication systems need to be represented and characterized. In our modeling framework, we represent the communication functions using additional queueing elements, hence yielding a queueing-network model for the holistic behavior of the PSAP (see Figure 2). Precisely, the acceptance of calls into the PSAP's gateway from landline and cellular networks is modeled as a fast multi-server queue with memoryless (exponential) servers, and with arrivals (service call times) defined by a time-varying Poisson or a batch Poisson process. The modeling of arrivals (service call times) as a Poisson process with time-varying rates has been justified in a number of empirical studies. The assumption that the call-acceptance service process is exponential is also often appropriate, however call acceptance does sometimes include processing services whose durations may have heavier-tailed distributions (for instance, locationing services may incur larger delays in some cases). The model can be extended to explicitly represent sequential or parallel processing during call acceptance with varying service-time distributions, however our initial simulations do not include this generalization. The accepted calls are then queued for handling by one of the available dispatchers, with the service by the dispatcher modeled as being memoryless (exponential). Again, more sophisticated service-time distributions can be assumed if needed, e.g. based on field-data studies [26,33].

When service by the dispatcher has been completed, then the dispatch process is initiated with some probability, representing that the call was judged to require service through dispatch. The dispatch process is represented using a hypercube model with multiple asset types (e.g., police, fire, ambulance), as developed in [29-32]. Specifically, as described above, the hypercube queueing model represents a geographical area that is served by the PSAP. Each call is modeled as originating from a stochastic location within the geographical area, with a uniform distribution used to model the location. Dispatch vehicles of multiple types are modeled as being originally located at stations with pre-specified locations within the geographical area. When a service call requires dispatch, the nearest salient, available resource is dispatched to the service location. The time required to reach the service location is modeled as a function of the distance, and thereafter the dispatch activity is modeled as requiring an exponentially-distributed amount of time. After the activity is completed, the resource is assumed to return to its station. During the dispatch process, the resource is assumed to be unavailable for other service.

The cyber-physical queueing-network model for a single PSAP is abstractly diagrammed in Figure 2 below.

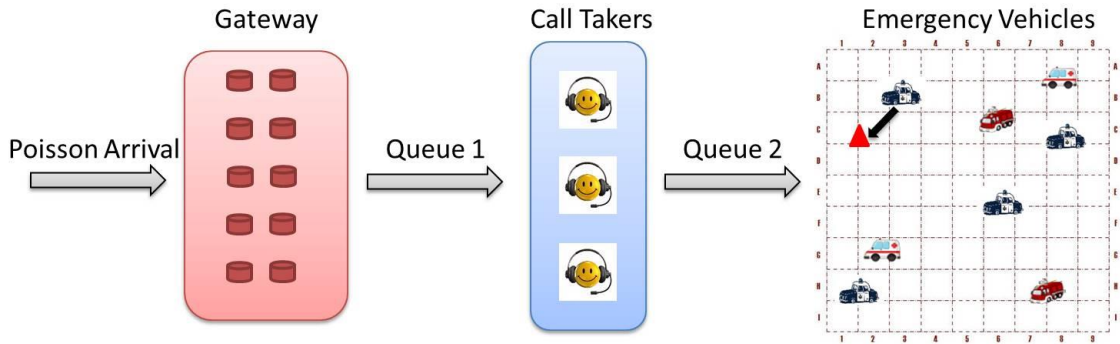

**Figure 2:** Cyber-physical queueing network model for a single PSAP.

The queueing-network model has also been enhanced to capture the additional access streams, modified IP-based architecture, and automated interconnectivity among PSAPs that are planned for NG911. Specifically, this has been done by first representing multiple access streams to an ESINet (e.g., landline, cell VoIO, text, image, and video) as distinct arrival streams, which may be subject to queueing at the front end of the ESInet. Second, the processing of client communications by the ESInet is represented using one or a network of queues with exponential servers; in this initial study, since architectures for ESInets are incompletely defined, we have used a simplified model representing the ESInet as a single exponential-server queue. Importantly, the queueing characteristics of the ESInet may have dependence on congestion in the Internet-at-large, and hence the queueing rate is allowed to be time-dependent. Third, the model captures that NG911 allows for automated routing and re-routing of calls among attendant PSAPs by the ESInet. Nominally, this function is simply modeled as routing each call based on its location information, to the PSAP responsible for the location. However, this nominal model may be further modified to capture usage-dependent routing, and also to include a protocol for routing when the location information is indeterminate. A reasonable representation for usage-dependent routing is that calls are rerouted if either the usage percentage of the PSAP call takers or of the dispatch resources exceeds a threshold. Fourth, for traditional PSAPs connected to an ESInet, a memoryless queueing element is used to represent the transfer of the call from the ESInet to the PSAP's gateway (i.e., from an IP network to a traditional-telephony gateway). The queueing model for NG911 is abstractly illustrated in Figure 3.

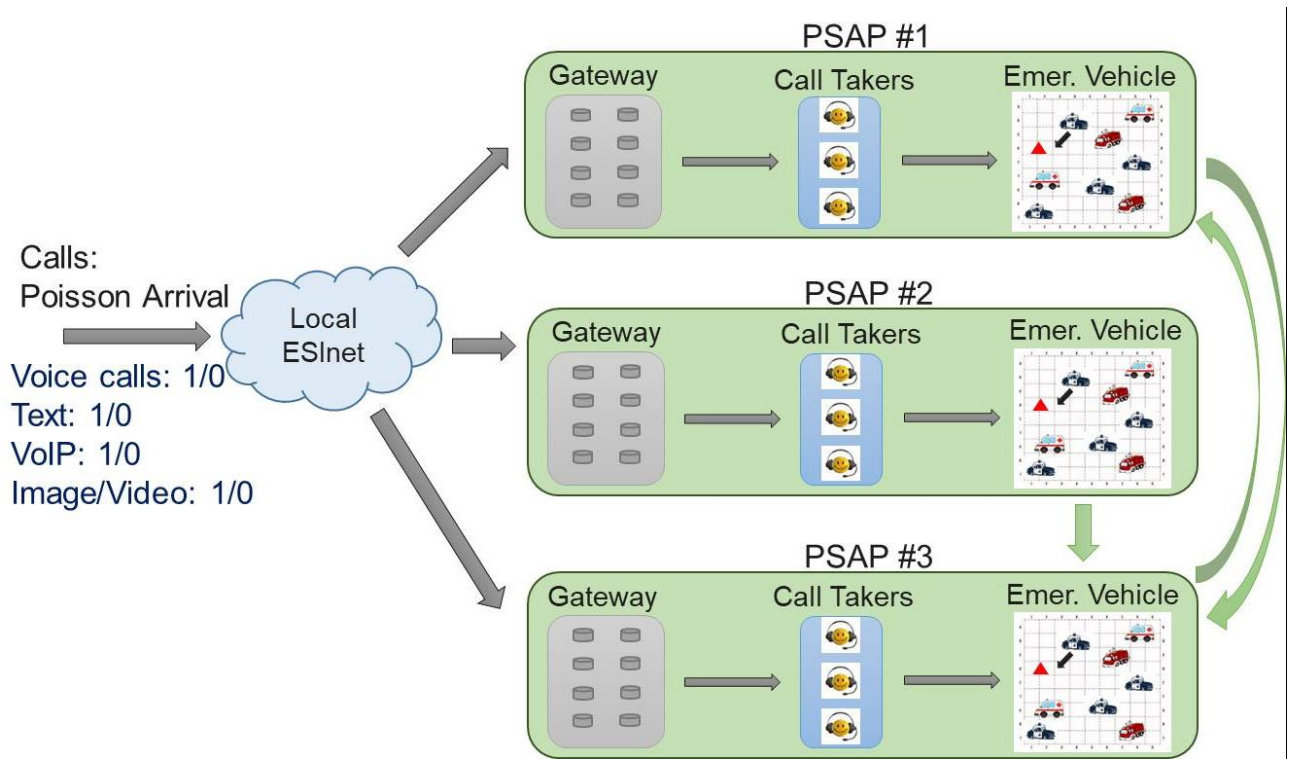

**Figure 3:** The extended end-to-end queueing network model for multiple PSAPs.

The described modeling framework is appealing for simulation/analysis of E911 and NG911 systems, for several reasons:

1) The parameters in the single-PSAP module can be obtained from real data. For example, call arrivals can be modeled as standard Poisson arrival processes. Historical call arrival times can be used to parameterize the Poisson arrival rates. Similarly, gateway slot service rates, call taker service rates, and emergency vehicle response times can also be simply obtained from historical data. For the multi-PSAP model for NG911, it is also expected that additional model parameters can be determined, as the technology choices for the NG911 system become more concrete. Even without full clarity about the NG911 architecture, the queueing-network description is sufficiently terse that queue parameters can be guessed, or the model can be tuned to match observed performance metrics.
2) The queueing-network model permits evaluation of standard performance metrics for the call-processing and dispatch functions (e.g. call-handling probabilities, call-processing times, total time until service). Further, the model allows interdependence analysis among such metrics: for instance, one can check whether calls with long total dispatch times also have long call-processing times, or whether call-processing times show a strong temporal correlation. Because the model captures communications traffic and dispatch at the flow level rather than at a detailed packet level, performance evaluation is also computationally attractive.

3) The model can potentially permit real time forecasting of transients and anomalies, in that future call-processing and dispatch statistics can be estimated given information about the current state.

## 4. Case Study: Envisioned NG911 System for Charlotte, NC

Modular software implementations of the queueing-network models for E911 and NG911 have been developed in Matlab. Specifically, we have developed a simulation of a study case, comprising an ESInet which is responsible for routing of calls to multiple PSAPs. The simulation encompasses the two main new features of the NG911 communication network -- specifically, the possibility for multiple communication options (standard landline and cell calls, text, Voice-over-IP, image/video transmission), and the possibility for call re-routing to different PSAPs based on geographical information or congestion. The developed simulation is modular, in the sense that it can naturally be extended to include new data streams and PSAPs. It can also be modified to be higher resolution, in that myriad services and processes envisioned in the NENA i3 architecture (e.g. GIS-based locationing services) can be included as further queueing elements. The simulation environment also allows assessment of vulnerabilities or risks in the NG911 architecture, including those resulting from the multifaceted access provided by the system, and those caused by the intertwining of the 911 system with the Internet-at-large (see Section 5).

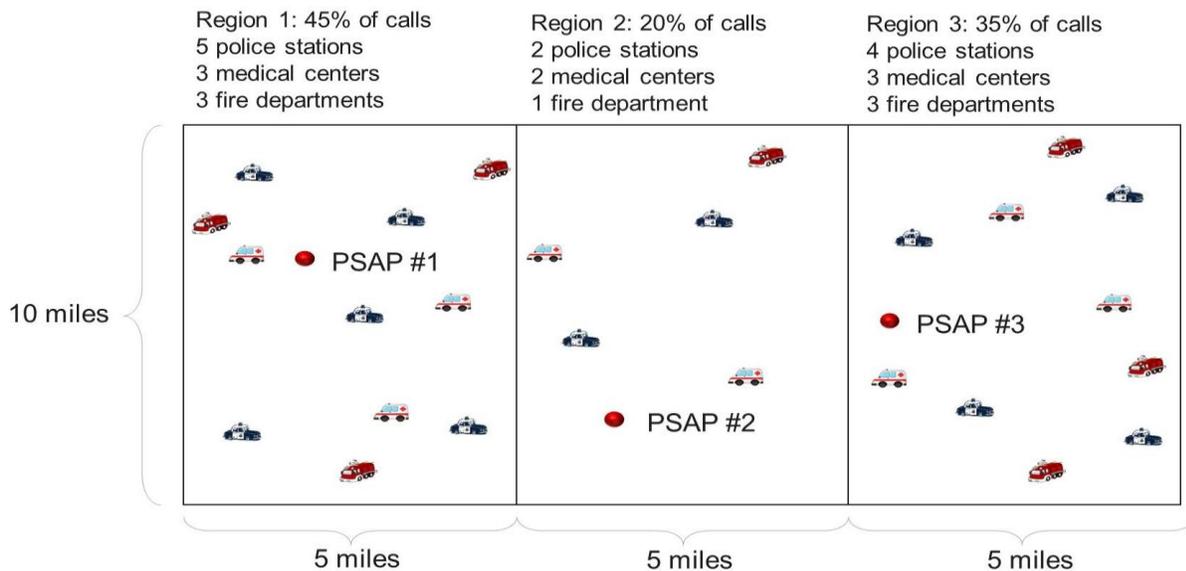

**Figure 4:** A three-region case study based on emergency services in the Charlotte, NC area.

The case-study instantiation of the model is roughly based on emergency services in a region in North Carolina (Charlotte area) served by three primary PSAPs. As a rough approximation, we consider a 15 mi by 10 mi jurisdiction (see Figure 4 above), which consists three regions with similar geographical size but different populations. While NG911 has not been brought into service in the Charlotte area, eventual updating of the 911 system to incorporate next-generation functions is anticipated. In envisioning a future NG911 architecture, an ESInet is defined that is connected to all three PSAPs. The parameters used in the multi-PSAP study case are as follows:

1) The simulation considers an 8-hour day-time span, with a fixed nominal call arrival rate at about 36 calls per hour. A higher call density is also considered for some simulations.
2) It is assumed that there are 100 routers in the local ESInet that can process 100 incoming calls simultaneously. Each router can process 50 calls per second on average. Of the total calls that use the local ESInet, on average 45%, 20%, and 35% of the total calls are routed to PSAP #1, #2, and #3, respectively. For each region, it is assumed that on average 75% of the received calls are voice, 10% are text messages, 10% are VoIP, and 5% are image/video messages.
3) Each PSAP's gateway contains 50 slots that can process 50 incoming calls simultaneously. It is assumed that each slot can process 100 calls per second on average.
4) Each PSAP has 3 call takers at all times. Each call taker spends about 2 minutes on average to process a voice call, 1 minute for text calls, 2 minutes for VoIP, and 1.5 minutes for image/video calls.
5) The total numbers of dispatch stations (i.e. police patrol divisions, comparable fire/ambulance) are 11, 5, and 10, respectively, and each station has two service vehicles. The average service time of a police car/ambulance/fire truck is 10/20/40 minutes.

The model parameters were estimated based on archived aggregate data from Charlotte NC [33], or from knowledge of typical emergency communications parameters.

Simulations of the model were undertaken, to understand whether the model provides reasonable statistical predictions of NG911 system performance metrics such as call processing times, total service time (time until dispatch unit reaches the emergency location), and call drop/abandonment rates. In addition, additional analyses permitted by the model were considered. In Figure 5, predictions of call-processing and service times are presented, to check whether the model yields predictions that are commensurate with field data. The call-processing and service time distributions and statistics are seen to be similar to distributions obtained from field data at several locations, including from Charlotte and from Montgomery County, Maryland [26,33,34].

The developed queueing model can be used in several ways, including for: a) understanding correlations between call processing and dispatch; b) comparing E911 and NG911 systems; c) studying workload; d) evaluating changing call demographics (e.g. growing use of text/image/video). As an illustration, Figure 5 presents call processing and total service times for each call to a PSAP in the model. This data allows analysis of the correlation between processing and dispatch, and in turn causative analysis of long service time events (i.e., whether call processing or dispatch is the cause).

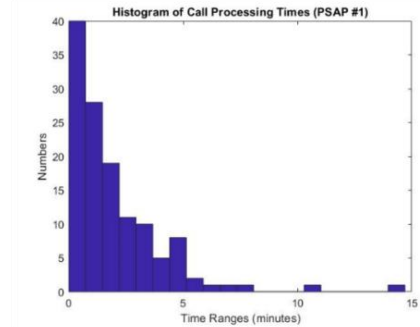
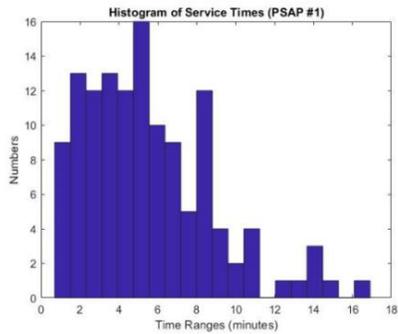

**Figure 5:** Call-processing time and total service time distributions (histograms) predicted by the model for the study case. The statistics and histogram are commensurate with distributions obtained from field data [26,33,34], suggesting that model is promising for end-to-end modeling of the 911 system.

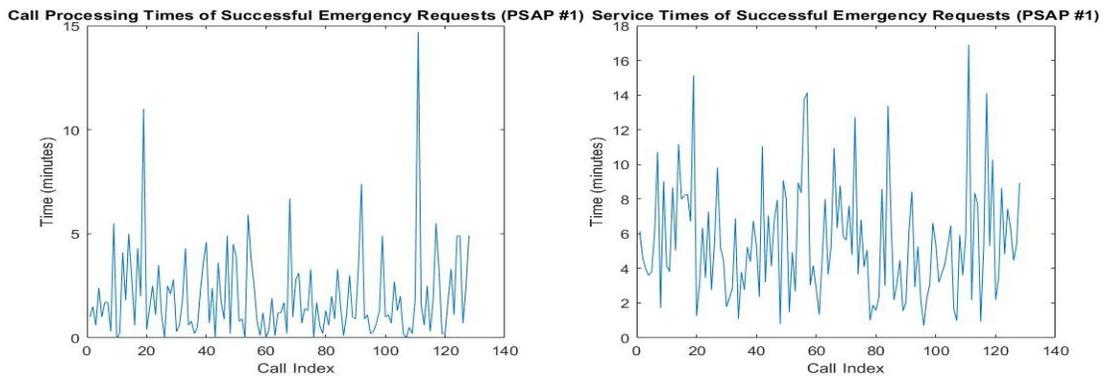

**Figure 6:** The model allows analysis of interactions between call-processing and total servce times. Here, call processing times and total service times of individual calls during a simulation run are shown side-by-side. It is seen that a significant fraction of calls with large service times also have anomously-large call-processing times.

## V. Threat Assessment Using the Cyber-Physical Queueing-Network Model

The literature on cyber-security for emergency response systems, including the Cyber-Security Primer for 911, identifies several classes of cyber- threats that are of significant concern [6,7]. These include device/operator-level attacks like malware and phishing; network-level attacks including Distributed Denial of Service (DDoS) attacks, Telephony Denial of Service (TDoS) attacks, and man-in-the-middle attacks; and application or service-layer attacks like swatting attacks and data breaches. Many of these threat classes are expected to have higher risk in NG911, both because their likelihood is higher given the multi-modal access to the system, and because the additional interconnectivity of the system may incur wider-area consequences. The additional complexity of the NG911 system is also likely to amplify the risks associated with failure events.

We have studied the representation of threats within the cyber-physical queueing-network model, and the consequent assessment of threat impacts through simulation of the model. Broadly, threats arising from the cyber-world can be modeled as modifying input or arrival processes (either to capture additional true service need or to represent falsified inputs), changing queue service rates or other characteristics, or changing routing protocols. Figure 7a describes how example threats can be represented in our queueing-network modeling framework. Also, in Figure 7b, the representation of DDoS attack in the queueing-network model is diagrammed as an overlay on the queueing-network model. Specifically, denial-of-service is incorporated into the model as additional incoming data streams (arrival processes) in the queueing-network model, which utilize server resources. We note that DDoS attacks are restricted to Internet-based systems, hence only apply to the NG911 system. Specifically, we model DDoS attacks as inserting data streams into the ESInet, which may either utilize server resources for a single PSAP or multiple PSAPs. Meanwhile, TDoS attacks may occur in either traditional 911 or NG911 systems, however more attack pathways are present in NG911 because the next-generation system has additional call options (e.g. VoIP).

| Threat | Modeling in Queueing Network: Examples |
|---|---|
| **Malware/Ransomware** | **Queue service failure (cyber)** |
| **Distributed Denial-of-service** | **Fake input process (cyber)** |
| **Swatting** | **Additional service need (cyber, dispatch)** |

a)

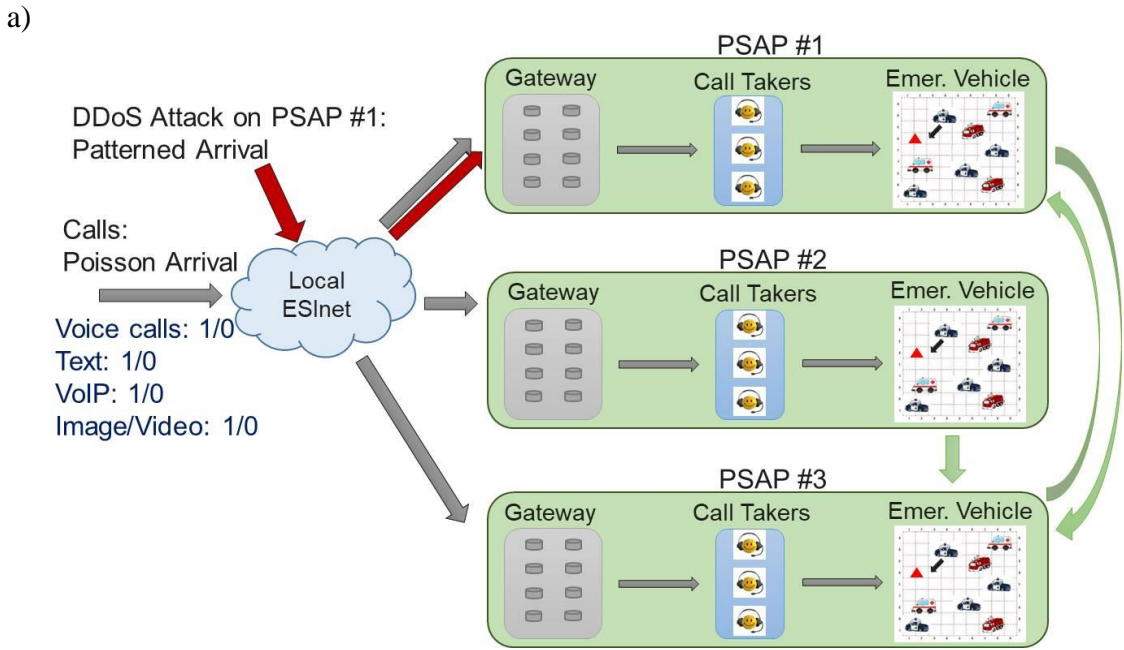

**Figure 7:** a) Representing threats in the queueing-network model: examples; b) Diagram of the queueing-network model with a DDoS attack included as an additional fake-data input.

Our interest is in characterizing holistic or mission-level impacts of threats such as the ones described above, as a step toward identifying key threats and computing risks. One key aspect of this analysis is defining metrics for holistic impact or consequence that are specialized for the emergency response system. For 911 systems, the primary concerns often are: 1) fidelity of the call and dispatch process when emergency services are needed; and 2) delay between call initiation and service at the dispatch location. Thus, natural impact metrics for an attack are: 1) additional call drops incurred by the attack, 2) call processing times, and 3) excess initiation-to-service delay caused by the attack. Because of the high variability of the emergency response system, statistics of the metrics (average, standard deviation) need to be computed. For the basic and enhanced 911 systems, these metric statistics can be computed for a single public safety access point (PSAP), since each PSAP operates independently. For NG911, metric statistics need to be computed both for individual PSAPs and for the broader region governed by an ESInet, since communications and possibly dispatch are coupled across the region.

Threat assessment has been undertaken for the case-study model derived from Charlotte, NC data. Example results for the impacts of DDoS attacks of different durations are shown in Figure 8. The attacks cause call drops/abandonment by tying up server resources, and also cause an increase in average processing and service times. The model quantifies and gives insight into these impacts – for instance, that a longer attack has a disproportionate impact on delays.

|  | No Attack | 5-min Attack | 30-min Attack |
| --- | --- | --- | --- |
| # of processed normal requests | 130 | 129 | 127 |
| # of dropped normal requests | 0 | 2 | 7 |
| Call drop/abandonment rate | 0% | 1.5% | 5.2% |
| Average call-processing time | 1.99 min | 2.65 min | 22.46 min |
| Average service time | 5.54 min | 6.40 min | 26.68 min |

**Figure 8:** Assessment of DDoS attacks of different durations using the model is shown, as an illustrative example. The model allows quantitative evaluation of threat consequences. Call-processing and service times are disproportionately impacted by longer-duration attacks.

## VI. Implications on Risk Assessment/Management and Future Work

Queueing-network models for traditional 911 systems as well as Next-Generation 911 systems have been introduced, which encompass the call-processing and dispatch functions of these systems. The queueing-network models are promising as tools for statistical analysis of 911 systems under nominal conditions, and also for assessment of the holistic impacts of cyber threats. The threat-impact analysis is a natural starting point for cyber- risk analysis of emergency response systems. In particular, we recall that risk metrics incorporate the likelihood and the consequence of threats. A likelihood analysis requires enumeration and analysis of threat modalities in the cyber- world at large, while the consequence analysis is specific to the emergency response system and is enabled by the developed models. Also of importance, the models can allow identification of high-impact threats as well as vulnerabilities, i.e. system components or processes that are particularly susceptible to attacks/failures that may have large impact. Thus, we believe that the developed queueing-network models are promising tools for risk and vulnerability assessment for E911 and NG911 systems.

We anticipate several directions of further work. First, we expect to develop detailed attributional descriptions of NG911 components which allow automated construction of queueing subsystems for the components (e.g., for an ESInet). Second, we anticipate extending the initial threat assessment effort toward a more comprehensive risk and vulnerability management study. Third, we expect to pursue formal analyses of the queueing-network models, with the goal of gaining some structural insights into the interdependencies between cyber- and physical- world components in 911 systems. Fourth, we anticipate using the models to study performance degradations and risks that may be introduced by interoperability issues in NG911. We also would like to draw on these methodological efforts to develop cyber- situational awareness tools for 911 operators, such as diagnostic tools and periodic e-mail reports which indicate emerging threats.